\documentclass[]{aastex701}
\usepackage{amsmath}
\graphicspath{{./}}
\shorttitle{The local galaxy distribution does not violate the cosmological principle}
\shortauthors{Till Sawala}

\providecommand{\apjl}{ApJ}
\providecommand{\apjs}{ApJS}
\providecommand{\ao}{Appl.~Opt.}

\providecommand{\aap}{A\&A}

\begin{document}

\title{The local galaxy distribution does not violate the cosmological principle}

\author[0000-0003-2403-5358]{Till Sawala}
\affiliation{University of Helsinki, Department of Physics, Helsinki, Finland}
\email[show]{till.sawala@helsinki.fi}

\begin{abstract}
The cosmological principle, which states that the Universe is statistically homogeneous and isotropic on sufficiently large scales, is a foundational assumption of the standard cosmological model. A recent analysis of DESI DR1 galaxy samples reported coherent anisotropic features in the local galaxy distribution extending to gigaparsec scales. If correct, this result would directly contradict the cosmological principle and motivate inhomogeneous cosmologies. Here I analyze the same data and compare them with galaxy distributions predicted by the FLAMINGO cosmological hydrodynamic simulation, performed in the standard $\Lambda$CDM paradigm. I show that the apparent anomaly disappears when the correct comoving distance scale is used and when the observations are compared to mock catalogs that account for bias and redshift-space distortions. Rather than violating the cosmological principle, the observed structures are consistent with those expected in a $\Lambda$CDM Universe.
\end{abstract}

\section{Introduction}\label{sec:intro}
The standard paradigm of hierarchical structure formation predicts a Universe that is strongly inhomogeneous on the scales of galaxies and galaxy clusters, but statistically homogeneous and isotropic on sufficiently large scales \citep{Peebles-1980,Davis-1985}. Individual, extended structures have been observed since the first redshift surveys \citep{Huchra-1982,DeLapparent-1986,Geller-1989,Gott-2005}, and modern observations reveal a complex cosmic web of voids, filaments, clusters and superclusters \citep[e.g.][]{Tully-2014,Valade-2024}. Such features are also a natural outcome of cold-dark-matter simulations \citep{White-1987a,Springel-2005-millennium,Park-2012,Sawala-2024,Schaller-2024},
and indeed required to explain the observed distribution of structures in the Local Universe.

There have been several claims about discoveries of stronger, or even larger-scale clustering than predicted by the standard model. These include local underdensities \citep{Keenan-2013}, source-count dipoles \citep{Secrest-2021,LandStrykowski-2025}, anisotropies in galaxy-cluster scaling relations \citep{Migkas-2020}, supernovae \citep{Sorrenti-2023}, peculiar velocities \citep{Watkins-2023, Courtois-2025}, or gigaparsec-scale patterns \citep{Lopez-2022}.

In general, sufficiently strong departures from homogeneity or isotropy would call into question the applicability of the standard Friedmann--Lema\^itre--Robertson--Walker (FLRW) framework \citep[e.g.][]{Buchert-2012, Lapi-2023, Constantin-2023}. Individually or collectively, observations of extremely large structures and other large-scale anomalies have also been interpreted as evidence for a breakdown of the cosmological principle or motivation for non-standard physics \citep[e.g.][]{Aluri-2023, Lopez-2024b, Mazurenko-2025, Meissner-2025}. However, sparse sampling, survey geometry, a posteriori feature selection and look-elsewhere effects can also produce individual apparently significant patterns. Several claimed structures were later shown to be statistically insignificant, consistent with random fluctuations, or compatible with standard model predictions \citep{Nadathur-2013, Park-2015, Marinello-2016, Balazs-2018, Fujii-2024, Sawala-2025}.

A recent work by \cite{SylosLabini-2026d} makes a broader, and apparently stronger claim: that the DESI (Dark Energy Spectroscopic Instrument) DR1 (Data Release~1) galaxy distribution contains statistically significant anisotropic structure on gigaparsec scales. Using a volume-limited sample from the bright galaxy catalog of DESI DR1 \citep{Hahn-2023, AbdulKarim-2026}, they report galaxy clustering far in excess of geometry-matched mocks based on $\Lambda$CDM simulations.

By including a large fraction of the public DESI DR1 survey data, this analysis avoids any suspicion of cherry-picking that could give rise to a look-elsewhere effect. The comparison to a large sample of simulation mocks also appears to provide a robust and even conservative demonstration of statistical significance. A detection of clustering at this magnitude and statistical significance would thus directly challenge and possibly falsify the cosmological principle, that is, the statistical homogeneity and isotropy inherent not only in $\Lambda$CDM but in the broader FLRW framework. The present work examines the same DESI DR1 data, but arrives at the opposite conclusion.

\section{The DESI DR1 data}
Following \citet{SylosLabini-2026d}, the observational data used here consist of the
public DESI DR1 spectroscopic galaxy catalog \citep{AbdulKarim-2026}, and in
particular the Bright Galaxy Sample \citep[BGS;][]{Hahn-2023}. The BGS is a low-redshift,
high-density DESI sample selected from Legacy Surveys photometry. In their work,
\citet{SylosLabini-2026d} restricted the analysis to a contiguous, nearly uniform subregion.
Their main BGS sample, labeled \texttt{S2}, contains $N=36{,}290$ galaxies with $M_r<-20$
within a cylinder of reported radius $R=300\,h^{-1}\,\mathrm{Mpc}$ and thickness
$40\,h^{-1}\,\mathrm{Mpc}$.

The present work is based on the same public BGS catalog, which is complete to $M_r<-21.5$ to the maximum distance considered. For the full $R = 290\,h^{-1}\,\mathrm{Mpc}$ comparison,  $M_r<-21.55$ is used to match the surface number density of the $\texttt{S2}$ sample in its fiducial coordinates, while galaxies up to $M_r<-20$ are included in smaller subvolumes. Importantly, comoving distances are computed directly from the published spectroscopic redshifts. Throughout this work, the D3A cosmology is adopted, with $h=0.681$, $\Omega_{\rm m}=0.306$, $\Omega_{\rm b}=0.0486$, $\Omega_\Lambda=0.694$, $\sigma_8=0.807$, $n_{\rm s}=0.967$, and $\sum m_\nu=0.06\,{\rm eV}$. Details are given in Appendix~\ref{sec:methods:DESI}.

\begin{figure*}
\centering 
\includegraphics[width=18cm, trim={0.0cm 0.0cm 0.0cm 0.0cm},clip]{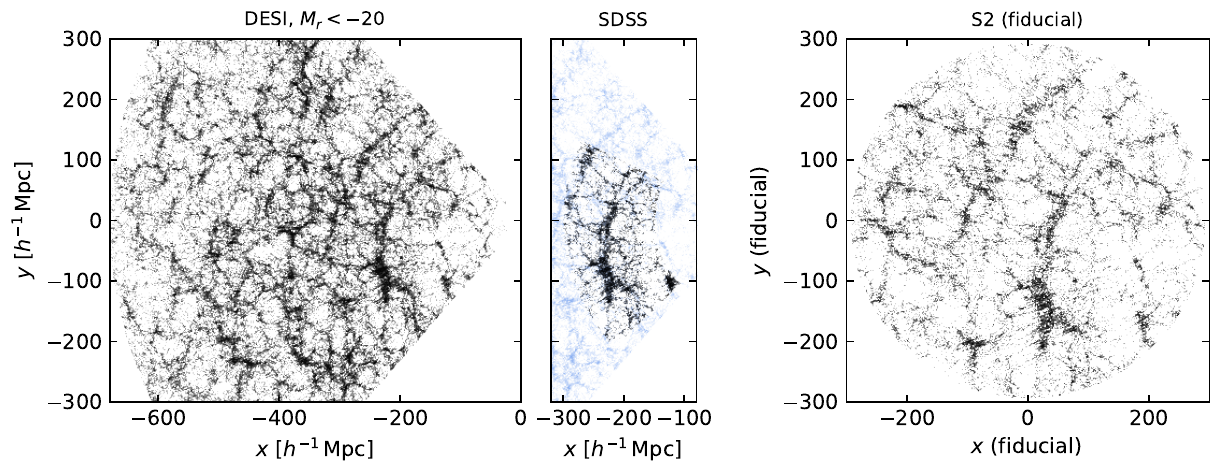}
\caption{\label{fig:DESI-SDSS-S2} Comparison of galaxy distributions in DESI DR1, SDSS, and \texttt{S2}. The left panel shows the distribution of galaxies in a slice of thickness $40\,h^{-1}\,\mathrm{Mpc}$ within the DESI DR1 footprint. The central panel shows galaxies in the overlap region with SDSS centered on the ``Sloan Great Wall'', with DESI data shown in light blue, and SDSS data shown in black. The observer is located on the right, at $x=0$. The right panel shows the spatial distribution in the \texttt{S2} data, with a diameter of 600, and the observer assumed at $x \simeq 400$, both in the published \texttt{S2} coordinates. The prominent structure visible in \texttt{S2} can be identified as an enlarged and slightly distorted projection of the Sloan Great Wall.}
\end{figure*}

The left panel of Figure~\ref{fig:DESI-SDSS-S2} shows the distribution of galaxies within the DESI DR1 footprint. The position of the observer is evident from the geometry of the cone and the concentric redshift-space distortions. The DESI data extend to $680\,h^{-1}\,\mathrm{Mpc}$ and enclose a cylinder of $R=290\,h^{-1}\,\mathrm{Mpc}$ with a center at a distance of $390\,h^{-1}\,\mathrm{Mpc}$, corresponding to redshift $z=0.13$, and spanning a redshift range of $z \simeq 0.03-0.24$.

The central panel of Figure~\ref{fig:DESI-SDSS-S2} shows the overlap between DESI DR1 and the Sloan Digital Sky Survey Data Release 18 \citep{Almeida-2023}, in a region defined by \cite{Einasto-2011-SGW}. The region is centered on the ``Sloan Great Wall", a prominent structure in the Local Universe \citep{Gott-2005} at a comoving distance of $\sim 235\,h^{-1}\,\mathrm{Mpc}$. The right panel showing the \texttt{S2} data is discussed in Section~\ref{sec:explanation}.

\section{The FLAMINGO simulation}
To test whether the structures in DESI DR1 are compatible with those expected in the standard model, geometry-matched mock galaxy catalogs are constructed based on the FLAMINGO cosmological hydrodynamic simulation \citep{Schaye-2023,Kugel-2023,Helly-2026}. The particular \texttt{L1\_m8} run evolves a periodic cube of side length $L=1000\,\mathrm{cMpc}$ containing $3600^3$ baryonic particles, $3600^3$ CDM particles and $2000^3$ neutrino particles, corresponding to initial particle masses $m_{\rm gas}=1.34\times10^8\,{\rm M_\odot}$ and $m_{\rm CDM}=7.06\times10^8\,{\rm M_\odot}$. The simulation also adopts the D3A cosmology. The galaxy-formation model includes radiative cooling, star formation, stellar mass loss and chemical enrichment, supernova feedback, massive black holes and thermal AGN feedback. Halos and galaxies are identified with HBT-HERONS \citep{Han-2018,ForouharMoreno-2025}, and their properties are computed with SOAP \citep{McGibbon-2025}. Importantly, FLAMINGO permits the construction of galaxy mocks which capture the effects of halo bias \citep{Kaiser-1984} and galaxy bias \citep{Kauffmann-1997}, in addition to redshift-space distortions \citep{Kaiser-1987}, all of which modify the measured clustering strength relative to the real-space density field, and are required for a fair comparison to observations. To construct the mock samples, galaxies are ranked by $r$-band luminosity, and simulated redshift-space distortions are added according to the peculiar velocity data and an observer at the same position relative to the DESI field. Details are given in Appendix~\ref{sec:methods:FLAMINGO} and Appendix~\ref{sec:methods:RSD}.

\begin{figure*}
\centering 
\includegraphics[width=18cm, trim={0.0cm 0.0cm 0.0cm 0.0cm},clip]{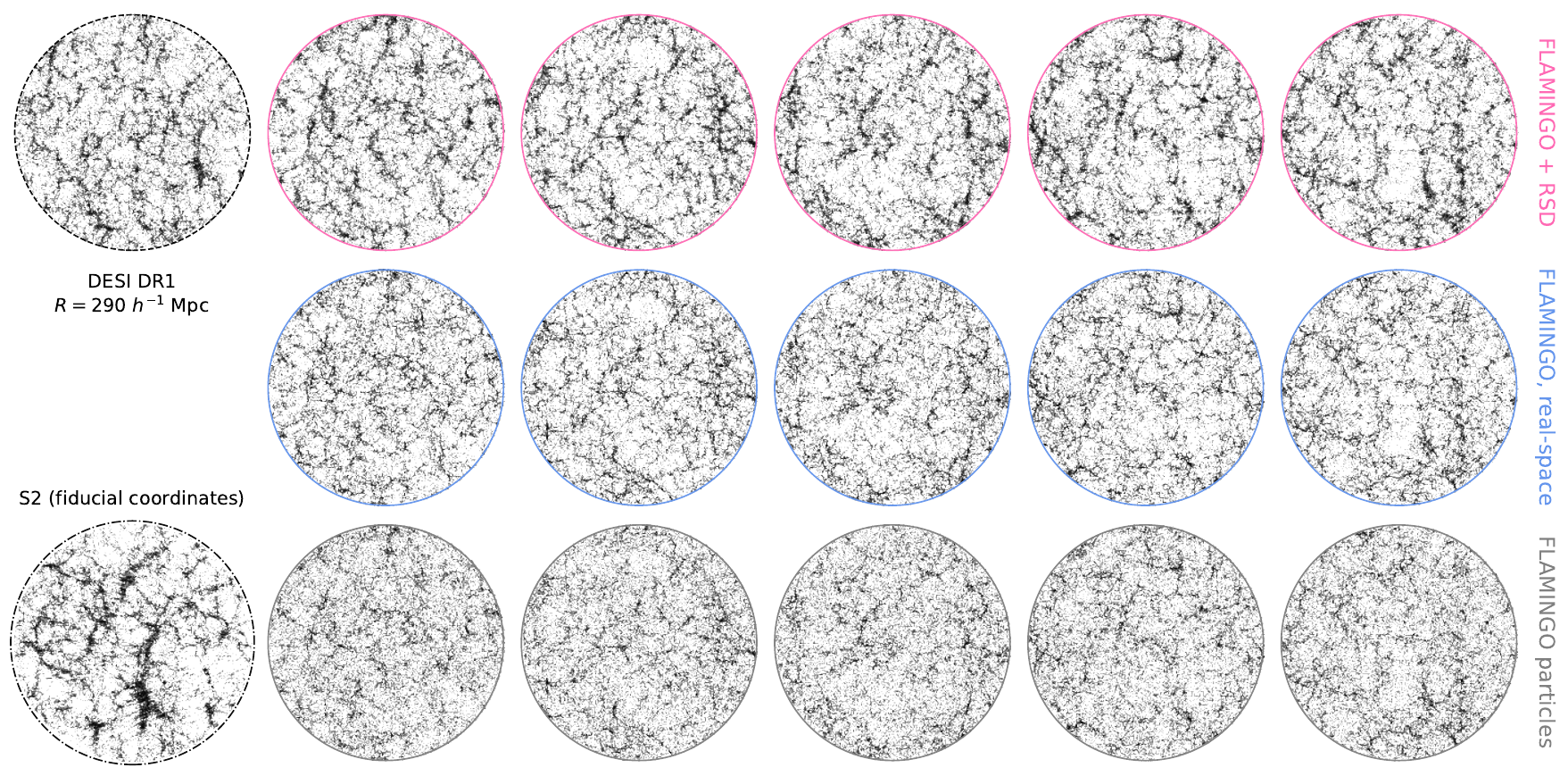}
\caption{\label{fig:structure-comparison-large} Distribution of galaxies from DESI DR1, \texttt{S2}, and in five non-overlapping geometry-matched cylinders selected from FLAMINGO at $z=0.15$. The top-left panel shows galaxies within an $R = 290\,h^{-1}\,\mathrm{Mpc}$ cylinder in DESI, the next five panels show galaxy positions from FLAMINGO including redshift-space distortions. The middle row shows FLAMINGO galaxies in real space without redshift-space distortions. The bottom-left panel shows galaxies in \texttt{S2} at its fiducial scale, the next five panels show the FLAMINGO density field sampled by dark-matter particles. The amount of structure in DESI DR1 is visually similar to that in FLAMINGO galaxy catalogs with redshift-space distortions included. Without redshift-space distortions, the clustering is weaker and the clustering in the particle distribution is the weakest. By contrast, at its fiducial scale, \texttt{S2} shows a very strong excess of structure, especially compared to particle-based mocks.}
\end{figure*}

\section{Clustering in DESI DR1 compared to $\Lambda$CDM predictions}
Figure~\ref{fig:structure-comparison-large} provides a visual comparison of the projected galaxy distribution in DESI to those in five non-overlapping simulation regions (stacked along the simulation $z$-axis) of radius $R = 290\,h^{-1}\,\mathrm{Mpc}$ at $z=0.15$. The amount of structure in DESI DR1 visually resembles that in the mock galaxy catalogs of the FLAMINGO simulation, once redshift-space distortions are included. The absence of redshift-space distortions reduces the clustering strength relative to realistic galaxy mocks, and the direct sampling of simulation particles further reduces the amount of clustering. Compared at its fiducial scale, the clustering in the \texttt{S2} data is far stronger than in the simulation.

\begin{figure*}
\centering 
\includegraphics[width=8.5cm, trim={0.0cm 0.0cm 0.0cm 0.0cm},clip]{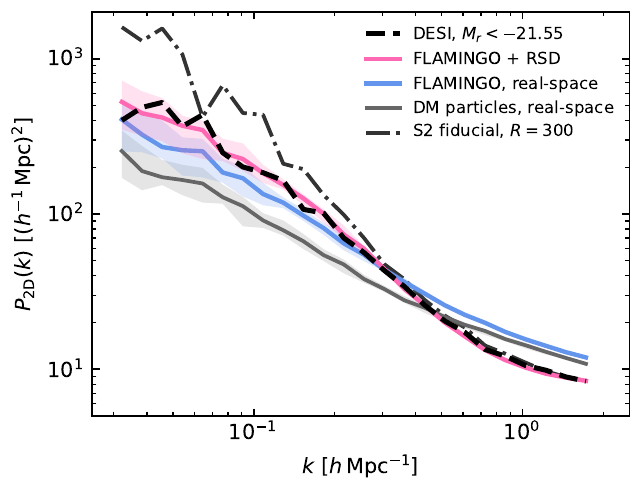}
\caption{\label{fig:DESI-power-spectra}Power spectra comparison between DESI, FLAMINGO, and \texttt{S2}. The dashed black line shows the 2D windowed power spectrum in the $R = 290\,h^{-1}\,\mathrm{Mpc}$ cylinder from DESI DR1, the dash-dotted line shows the power spectrum measured for \texttt{S2} in fiducial coordinates. Colored lines show the median power spectra in geometry-matched slices of the FLAMINGO simulation at $z=0.15$, measured using galaxies accounting for redshift-space distortions (pink), galaxies in real space (blue), or simulation particles (gray). Bands show percentiles equivalent to $\pm 1 \sigma$ across all samples. The large-scale modes show considerable cosmic variance. The small-scale behavior is affected by both redshift-space distortions and shot noise. However, when redshift-space distortions are included, the DESI DR1 data are compatible with the FLAMINGO data on all scales. By contrast, the power spectrum measured for \texttt{S2} in fiducial coordinates would imply a highly significant tension with the model assumed in the simulation.
}
\end{figure*}

At first glance, the structures in the DESI slice appear very similar to those in the simulation. This visual impression is confirmed by a comparison of the power spectra in Figure~\ref{fig:DESI-power-spectra} (see Appendix~\ref{sec:methods:PS} for details). The results from the FLAMINGO simulations are presented for galaxy catalogs in redshift space as well as in real space, and for simulation particles. The effect of galaxy- and halo bias boosts the clustering of real-space galaxy positions relative to the density field sampled by dark matter particles, particularly on large scales. The effect of redshift-space distortions is a boost in power on large scales and a strong suppression on small scales. Taking into account the cosmic variance affecting the large-scale modes, the simulation is in remarkably good agreement with the DESI data on all scales. The clustering strength of the DESI DR1 data is in excellent agreement with the prediction of the standard cosmological model when compared to a state-of-the-art simulation like FLAMINGO.

\begin{figure*}
\centering 
\includegraphics[width=18cm, trim={0.0cm 0.0cm 0.0cm 0.0cm},clip]{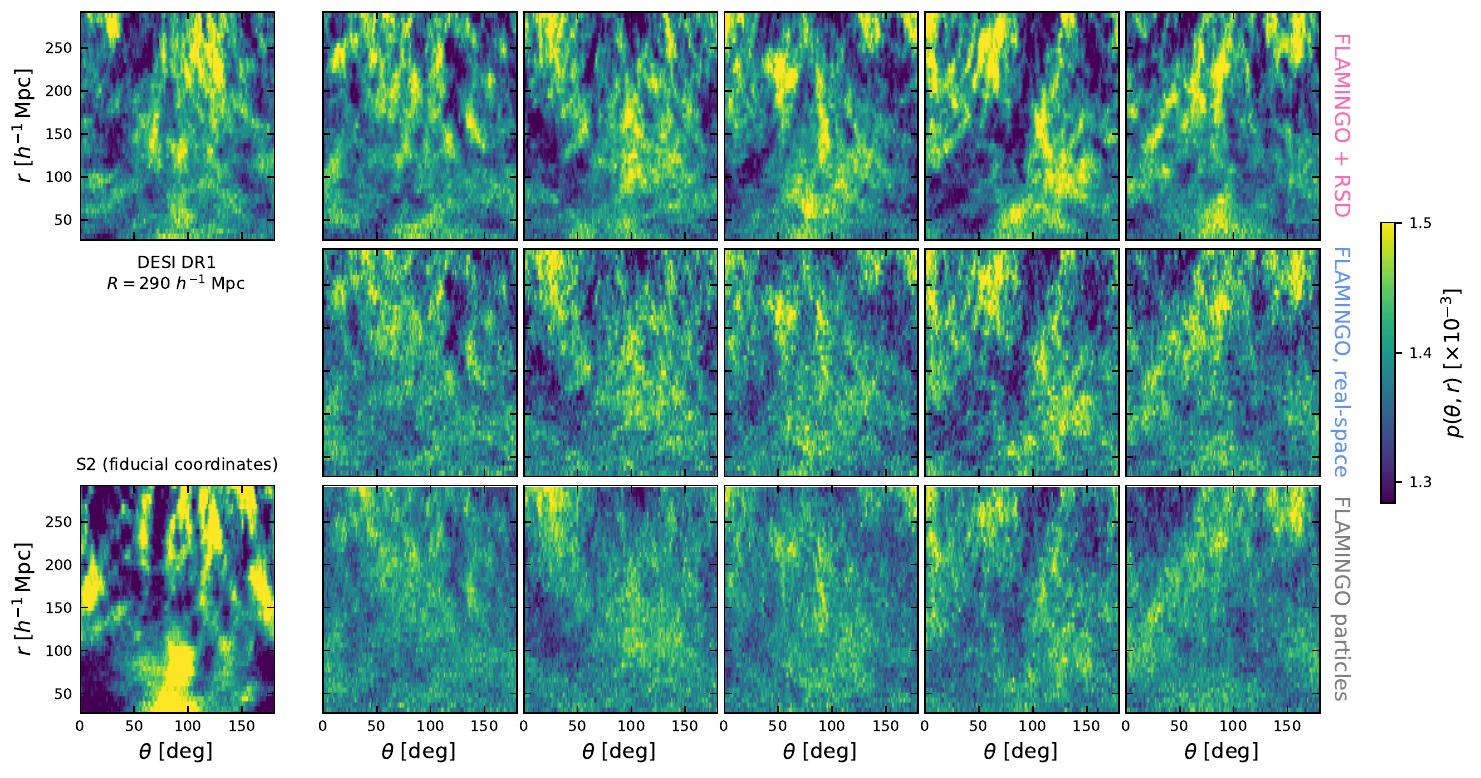}
\caption{\label{fig:ADPD} Anisotropy, measured as the angular distribution of pairwise distances, as a function of pairwise distance, $r$, and angle, $\theta$. The top left panel shows the result within an $R = 290\,h^{-1}\,\mathrm{Mpc}$ cylinder in DESI, the bottom left panel shows the result for \texttt{S2} at its fiducial scale. The next five columns show the results from FLAMINGO in the same five volumes shown in Figure~\ref{fig:structure-comparison-large}. The top row shows results for FLAMINGO galaxies including redshift-space distortions, the middle row shows FLAMINGO galaxies in real space, the bottom row shows results measured from the dark-matter particles. FLAMINGO reproduces the statistical anisotropy observed in DESI, but only when compared at the correct scale, and when bias and redshift-space distortions are included. The anisotropy that would be inferred in \texttt{S2} is much greater.}
\end{figure*}

To quantify the anisotropy, the same angular distribution of pairwise distances as used in \citet{SylosLabini-2026d} is computed for the DESI DR1 data, the FLAMINGO simulation, and for the \texttt{S2} data in its published coordinates (see Appendix~\ref{sec:methods:adpd} for details). Figure~\ref{fig:ADPD} shows the measured probability density, $p(\theta, r)$, as a function of separation, $r$, and angle, $\theta$. The variance of the probability density as a function of $\theta$, is a measure of anisotropy at a given scale $r$. From Figure~\ref{fig:ADPD} it can be seen that the FLAMINGO simulation reproduces the statistical anisotropy observed in DESI, if and only if compared at the correct scale, and when both bias and redshift-space distortions are included.

These conclusions stand in stark contrast with the findings of \citet{SylosLabini-2026d}. The dash-dotted black line in Figure~\ref{fig:DESI-power-spectra} shows the two-dimensional power spectrum measured from \texttt{S2} in its published coordinates. Consistent with the excess angular power reported in \citet{SylosLabini-2026d}, there is a very strong and highly significant excess of large-scale clustering, which would make the observations inconsistent with the standard model. 

\section{Explaining the Discrepancy} \label{sec:explanation}
The discrepancy between the results presented here and those of \citet{SylosLabini-2026d} is plausibly explained by a distance error inherited from the construction of the \texttt{S2} coordinates in \citet{SylosLabini-2026a}. The published \texttt{S2} coordinates appear to treat luminosity distances expressed in $\mathrm{Mpc}$ as if they were comoving distances in $h^{-1}\,\mathrm{Mpc}$, resulting in a spurious enlargement by a factor $\frac{1+z}{h}$. This has substantial implications for the inferred inhomogeneity and anisotropy. Details are provided in Appendix~\ref{sec:methods:scale}.

Figure~\ref{fig:distance-error} demonstrates the consequence of the distance error. The left panel shows the \texttt{S2} data in fiducial coordinates. The middle panel shows the DESI data, but in coordinates where the luminosity distance in Mpc is deliberately equated to comoving coordinates in $h^{-1}{\rm Mpc}$. It is worth noting that I do not compute independent luminosity distances, but simply apply the $\frac{1+z}{h}$ factor to the redshift-derived comoving distances. Nevertheless, this erroneous projection closely aligns with the \texttt{S2} data in fiducial coordinates. 

\begin{figure*}
\centering 
\includegraphics[width=18cm, trim={0.0cm 0.0cm 0.0cm 0.0cm},clip]{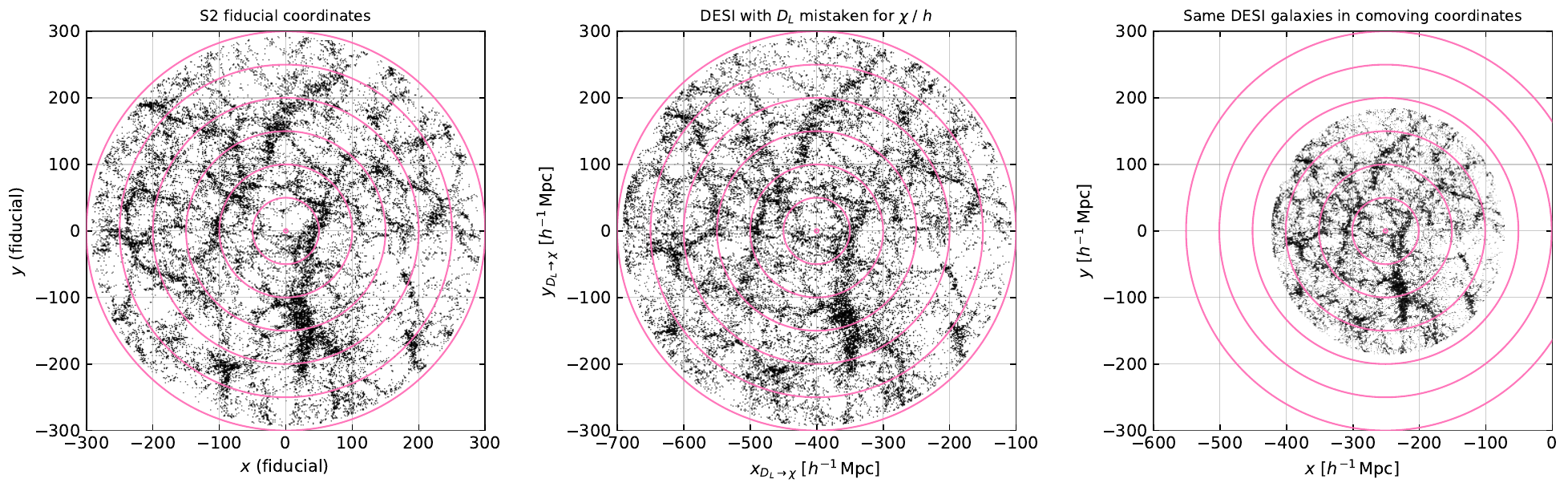}
\caption{\label{fig:distance-error} Consequence of the mistaken distance coordinates in the \texttt{S2} data set. The left panel shows the \texttt{S2} data in published coordinates. The center panel shows DESI data, but plotted using the mistaken identification of luminosity distance in $\mathrm{Mpc}$ with comoving distance in $h^{-1}\,\mathrm{Mpc}$. The right panel shows the same DESI data, but in true comoving coordinates. Pink circles are equally spaced in intervals of 50 units on each panel. The mistaken transformation both spuriously scales and deforms the data. Small differences in the galaxy sample arise from differences in the selection function, luminosity distance calculation, and assumed cosmology.
}
\end{figure*}

The right panel shows the same DESI galaxies selected from the same region that appears circular when equating luminosity distances to comoving distances, but projected in actual comoving coordinates. Comparison to circles placed at constant intervals in comoving coordinates demonstrates both the artificial scaling and the distortion brought about by the redshift-dependence.

An immediate consequence is also visible in Figure~\ref{fig:DESI-SDSS-S2}, which compares the DESI and SDSS data to the \texttt{S2} data, shown in the published coordinates, as presented in Figure~1 of \citet{SylosLabini-2026d}. The prominent structure close to the center of \texttt{S2} can be identified as the Sloan Great Wall, albeit slightly distorted and significantly enlarged. 

Due to the scale-dependent nature of clustering in the standard model evident in Figure~\ref{fig:DESI-power-spectra}, comparing structures at different scales can result in apparent discrepancies. In addition to boosting the apparent large-scale power, the distortion also introduces spurious anisotropy.

Furthermore, the comparison in \cite{SylosLabini-2026d} was made to simple dark matter particle mocks, equivalent to those shown in the bottom row of Figure~\ref{fig:structure-comparison-large}, whose power spectra are shown as gray lines in Figure~\ref{fig:DESI-power-spectra}. These only sample the matter distribution and significantly underestimate the amount of inhomogeneity and anisotropy in the galaxy distribution, already predicted by the standard model.

\section{Discussion}
In the hierarchical paradigm, clustering strength is strongly scale-dependent, making comparisons between observations and predictions highly sensitive to systematic distance errors. In addition, realistic predictions of the inhomogeneity and anisotropy of the galaxy distribution require mock galaxy catalogs that account for halo bias, galaxy bias, and redshift-space distortions.

Comparing DESI DR1 data to realistic mock galaxy catalogs based on equivalent regions in the FLAMINGO simulation, I have shown that the observed local large-scale structure is in excellent agreement with theoretical predictions in terms of both inhomogeneity and anisotropy. The earlier claim of a gigaparsec-scale departure from homogeneity and isotropy instead originates from an incorrect distance scale, as well as inadequate mock catalogs.

It is worth emphasizing that the absence of large-scale inhomogeneity or anisotropy is no proof that the standard model is correct: indeed, analyses by the DESI team, combining DESI BAO measurements with CMB and supernova data, show a small preference for a time-dependent dark-energy equation of state in the $w_0w_a$CDM model over $\Lambda$CDM \citep{DESI-DR2-2025-Cosmology, DESI-2024-VI-Cosmology}. However, based on the available data, there is no evidence for a violation of the cosmological principle.

\section*{Code}
The numerical analysis was performed in Python. Data processing and cosmological distances were computed using \texttt{astropy} \citep{Astropy-2013, Astropy-2022}, and the calculations including the Fourier transforms used \texttt{numpy} \citep{numpy-paper}. All figures were produced with \texttt{matplotlib} \citep{matplotlib-paper}. The anisotropy analysis was performed using code by \citeauthor{SylosLabini-2026d}. A notebook to perform the analysis and reproduce all figures in this paper from public data is provided at \url{https://github.com/TillSawala/DESI-FLAMINGO} and is
permanently archived on Zenodo at
\dataset[doi:10.5281/zenodo.21445576]
{https://doi.org/10.5281/zenodo.21445576}. An LLM (ChatGPT 5.5) was used to help develop the notebook.

\section*{Data}
All data used in this work are publicly available. The FLAMINGO data are available at \url{https://dataweb.cosma.dur.ac.uk:8443/flamingo/introduction.html}. The DESI DR1 data are available at \url{https://data.desi.lbl.gov/doc/releases/dr1/}. The SDSS data are available at \url{https://skyserver.sdss.org/dr18}. The \texttt{S2} data are available at \dataset[doi:10.5281/zenodo.20118015]{https://doi.org/10.5281/zenodo.20118015}.

\section*{Acknowledgments}
I thank Carlos Frenk, Peter Johansson, Gabor Racz and Syksy R\"as\"anen for very helpful discussions and comments. I thank Peter Coles and Seshadri Nadathur for additional helpful comments on the mock catalogs, and I thank Francesco Sylos Labini and Marco Galoppo for a helpful and constructive discussion, and for making their code and data publicly available. I also thank the anonymous referee for their prompt and encouraging review.

In addition to the authors of the open-source software listed above, I gratefully acknowledge the Virgo Consortium and the creators of the FLAMINGO simulation, the DESI collaboration, and the SDSS collaboration for making their data publicly available. This work is supported by Research Council of Finland grants 354905 and 339127. This work used facilities hosted by the CSC -- IT Centre for Science, Finland.

The FLAMINGO simulations were performed using the Durham Memory Intensive system managed by the Institute for Computational Cosmology on behalf of the STFC DiRAC facility (\url{www.dirac.ac.uk}).

This research used data obtained with the Dark Energy Spectroscopic Instrument (DESI). DESI construction and operations is managed by the Lawrence Berkeley National Laboratory. This material is based upon work supported by the U.S. Department of Energy, Office of Science, Office of High-Energy Physics, under Contract No. DE–AC02–05CH11231, and by the National Energy Research Scientific Computing Center, a DOE Office of Science User Facility under the same contract. Additional support for DESI was provided by the U.S. National Science Foundation (NSF), Division of Astronomical Sciences under Contract No. AST-0950945 to the NSF’s National Optical-Infrared Astronomy Research Laboratory; the Science and Technology Facilities Council of the United Kingdom; the Gordon and Betty Moore Foundation; the Heising-Simons Foundation; the French Alternative Energies and Atomic Energy Commission (CEA); the National Council of Humanities, Science and Technology of Mexico (CONAHCYT); the Ministry of Science and Innovation of Spain (MICINN), and by the DESI Member Institutions: \url{www.desi.lbl.gov/collaborating-institutions}. The DESI collaboration is honored to be permitted to conduct scientific research on I’oligam Du’ag (Kitt Peak), a mountain with particular significance to the Tohono O’odham Nation. Any opinions, findings, and conclusions or recommendations expressed in this material are those of the authors and do not necessarily reflect the views of the U.S. National Science Foundation, the U.S. Department of Energy, or any of the listed funding agencies.

Funding for the Sloan Digital Sky Survey V has been provided by the Alfred P. Sloan Foundation, the Heising-Simons Foundation, the National Science Foundation, and the Participating Institutions. SDSS acknowledges support and resources from the Center for High-Performance Computing at the University of Utah. SDSS telescopes are located at Apache Point Observatory, funded by the Astrophysical Research Consortium and operated by New Mexico State University, and at Las Campanas Observatory, operated by the Carnegie Institution for Science. The SDSS web site is \url{www.sdss.org}. SDSS is managed by the Astrophysical Research Consortium for the Participating Institutions of the SDSS Collaboration, including the Carnegie Institution for Science, Chilean National Time Allocation Committee (CNTAC) ratified researchers, Caltech, the Gotham Participation Group, Harvard University, Heidelberg University, The Flatiron Institute, The Johns Hopkins University, L'Ecole polytechnique f\'{e}d\'{e}rale de Lausanne (EPFL), Leibniz-Institut f\"{u}r Astrophysik Potsdam (AIP), Max-Planck-Institut f\"{u}r Astronomie (MPIA Heidelberg), Max-Planck-Institut f\"{u}r Extraterrestrische Physik (MPE), Nanjing University, National Astronomical Observatories of China (NAOC), New Mexico State University, The Ohio State University, Pennsylvania State University, Smithsonian Astrophysical Observatory, Space Telescope Science Institute (STScI), the Stellar Astrophysics Participation Group, Universidad Nacional Aut\'{o}noma de M\'{e}xico, University of Arizona, University of Colorado Boulder, University of Illinois at Urbana-Champaign, University of Toronto, University of Utah, University of Virginia, Yale University, and Yunnan University.

\appendix{}

\section{Selection and processing of the DESI Data} \label{sec:methods:DESI}

The DESI DR1 data consist of the clustering catalogs of the Bright Galaxy Survey (BGS). From these, the sky positions, redshift, and dereddened Legacy Survey fluxes are used, namely the columns \texttt{RA}, \texttt{DEC}, \texttt{Z}, \texttt{flux\_g\_dered}, and \texttt{flux\_r\_dered}. Only galaxies with finite sky coordinates, positive redshifts, and positive finite $g$- and $r$-band dereddened fluxes are retained.

All DESI redshifts are converted directly to comoving distances using the same D3A cosmology as the FLAMINGO simulation used for the mocks, with $h=0.681$, $\Omega_{\rm m}=0.306$, $\Omega_{\rm b}=0.0486$. The redshifts naturally include peculiar velocities, making the distances subject to redshift-space distortions. The selection is restricted to $127^\circ~\leq~{\rm RA}~\leq~225^\circ, \ \ -7^\circ~\leq~{\rm Dec}~\leq~3^\circ$ and to a comoving distance of $\leq 680\,h^{-1}\,{\rm Mpc}$.

To compute completeness and select sets of the $N$ brightest galaxies, an approximate rest-frame $r$-band absolute magnitude is constructed directly from the dereddened DESI fluxes, applying the low-redshift $r$-band $K$-correction
$$
K_r(z,g-r) = z\left(2.5 - 1.5(g-r)\right),
$$
where $g-r=m_g-m_r$, using the $g$ and $r$-band fluxes in the catalog. The absolute magnitude used for the selection is then
\begin{equation*}
M_r = m_r - {\rm DM}(z) - K_r(z,g-r),
\end{equation*}
with
\begin{equation*}
{\rm DM}(z)=5\log_{10}\left(\frac{D_{\rm L}(z)}{\rm Mpc}\right)+25,
\end{equation*}
where $D_{\rm L}(z)$ is the luminosity distance evaluated in the same cosmology. The sample is volume-limited at approximately $M_r<-21.5$. For direct visual comparison with the published \texttt{S2} sample, plots use the $M_r<-20$ selection where appropriate. The scatter plots and power spectra are computed from the brightest galaxies within a cylinder, to either match the surface number density of \texttt{S2} in fiducial coordinates (which results in $M_r \lesssim-21.55$) or, for smaller volumes, to $M_r<-20$.

The sky coordinates and comoving distances are then converted to Cartesian coordinates, with the $y$-axis aligned with the negative RA direction, chosen to match the visual orientation of the
published \texttt{S2} coordinates, and the $z$-axis aligned with the DEC direction. The $x$-axis corresponds to RA$=0$, DEC $=0$ and increases toward the observer.

\section{Construction of mocks from FLAMINGO} \label{sec:methods:FLAMINGO}

Mock projected galaxy catalogs are constructed from the FLAMINGO hydrodynamical simulation with box size $L=1000\,{\rm cMpc}$ ($681 \,h^{-1}\,{\rm Mpc}$) and mass resolution level \texttt{m8}. In particular, the data are contained in the following SOAP-HBT catalogs at redshifts $z= 0.25$, $z= 0.15$ and $z= 0.05$, respectively: \\
\texttt{FLAMINGO/L1\_m8/L1\_m8/SOAP-HBT/halo\_properties\_0073.hdf5} \\
\texttt{FLAMINGO/L1\_m8/L1\_m8/SOAP-HBT/halo\_properties\_0075.hdf5} \\
\texttt{FLAMINGO/L1\_m8/L1\_m8/SOAP-HBT/halo\_properties\_0077.hdf5}

Only the quantities needed for the comparison are streamed from these files: the subhalo center of mass, \texttt{BoundSubhalo/CentreOfMass}, the subhalo center-of-mass peculiar velocity, \texttt{BoundSubhalo/CentreOfMassVelocity}, and the stellar luminosities \texttt{BoundSubhalo/StellarLuminosity}, from which the provided $r$-band luminosity is used.

The cylinders are centered on the middle of the FLAMINGO box in the $xy$ plane, and the simulation is tiled along the $z$-axis with non-overlapping cylinders. A slab thickness of $\Delta z=40\,h^{-1}\,{\rm Mpc}$ results in 17 non-overlapping slabs through the $1000\,{\rm Mpc}$ box. For each cylinder, the number of brightest galaxies required to match the comparison region is selected.

\section{Redshift-space distortions}\label{sec:methods:RSD}

Both real-space and redshift-space mocks are computed. To add redshift-space distortions, an observer at an appropriate location $\boldsymbol d_{\rm obs}$ needs to be defined for each volume. Galaxies are then displaced along the line of sight to the observer according to their peculiar velocities, $\boldsymbol v_i$, and positions, $\boldsymbol d_i$.

For each galaxy, the line-of-sight unit vector is
$$
\hat{\boldsymbol n}_i
=
\frac{\boldsymbol d_i-\boldsymbol d_{\rm obs}}
     {|\boldsymbol d_i-\boldsymbol d_{\rm obs}|}.
$$

Under the assumption of an observer at rest with respect to the background, the line-of-sight peculiar velocity is $v_{\rm LOS,i}=\boldsymbol v_i\cdot \hat{\boldsymbol n}_i$. The redshift-space position is then computed as
\begin{equation*}
{\boldsymbol s}_i =
{\boldsymbol d}_i +
\frac{v_{{\rm LOS},i}}{aH(a)}\hat{\boldsymbol n}_i ,
\end{equation*}
where $a$ and $H(a)$ are evaluated at the snapshot redshift.

To preserve the number density and selection, redshift-space distortions are applied before galaxies are selected in cylinders. The DESI redshift uncertainties of order $10\,{\rm km\,s^{-1}}$ \citep{Lan-2023} are negligible for the scales considered here. As shown in Figures~\ref{fig:structure-comparison-large} and~\ref{fig:DESI-power-spectra}, the inclusion of redshift-space distortions is crucial to reproduce the observed projected galaxy distribution.

\section{Origin and implications of the scale discrepancy} \label{sec:methods:scale}
In a spatially flat Friedmann--Lemaître--Robertson--Walker cosmology, the physical
line-of-sight comoving distance is
\begin{equation*}
\chi(z)
=
\frac{c}{H_0}
\int_0^z
\frac{{\rm d}z'}
{\sqrt{\Omega_{\rm m}(1+z')^3+\Omega_\Lambda}}
\label{eq:comoving}
\end{equation*}
while the physical luminosity distance, $d_L$, can be inferred from the distance modulus,
\begin{equation*}
M_i
=
m_i
-
5\log_{10}
\left(
\frac{d_L}{10\,{\rm pc}}
\right)
-
K_i(z),
\label{eq:absolute_mag}
\end{equation*}
where $i$ denotes the band and $K_i$ denotes the $K$-correction. \cite{SylosLabini-2026a} use this relation to compute the $r$-band luminosity distances for galaxies in the DESI DR1 BGS sample. However, these luminosity distances are then presented, and propagated to coordinates in \texttt{S2}, as comoving distances in units of $h^{-1}{\rm Mpc}$. The physical comoving distance is related to the luminosity distance via
\begin{equation*}
\chi(z) = \frac{d_L(z)}{1+z},
\label{eq:dL}
\end{equation*}
and in the usual coordinates
\begin{equation*}
\chi(z) [h^{-1}{\rm Mpc}] = \frac{h}{1+z}d_L(z) [\rm Mpc].
\label{eq:dL2}
\end{equation*}

At $z=0$, with $h=0.681$, the spurious enlargement factor in \texttt{S2}, $\frac{1+z}{h}$, is 1.47. At the distance of the Sloan Great Wall ($z \sim 0.08$, $\chi \simeq 235 h^{-1}{\rm Mpc}$), it rises to 1.59, while at the outer edge of the DESI DR1 region ($z\simeq0.24$, $\chi \simeq 680 h^{-1}{\rm Mpc}$), the factor reaches 1.82.

For the \texttt{S2} region, the erroneous enlargement factor varies between $\sim 1.50$ and $\sim 1.68$. With nearby distances enhanced less than more distant ones, the 600~Mpc extent in luminosity distance along the line of sight corresponds to only $\sim 350 h^{-1}{\rm Mpc}$ in comoving distance.

The same coordinate error appears to have been made for the other data in \cite{SylosLabini-2026a}, possibly including the LRG sample.

\begin{figure*}
\centering 
\includegraphics[width=15cm, trim={0.0cm 0.0cm 0.0cm 0.0cm},clip]{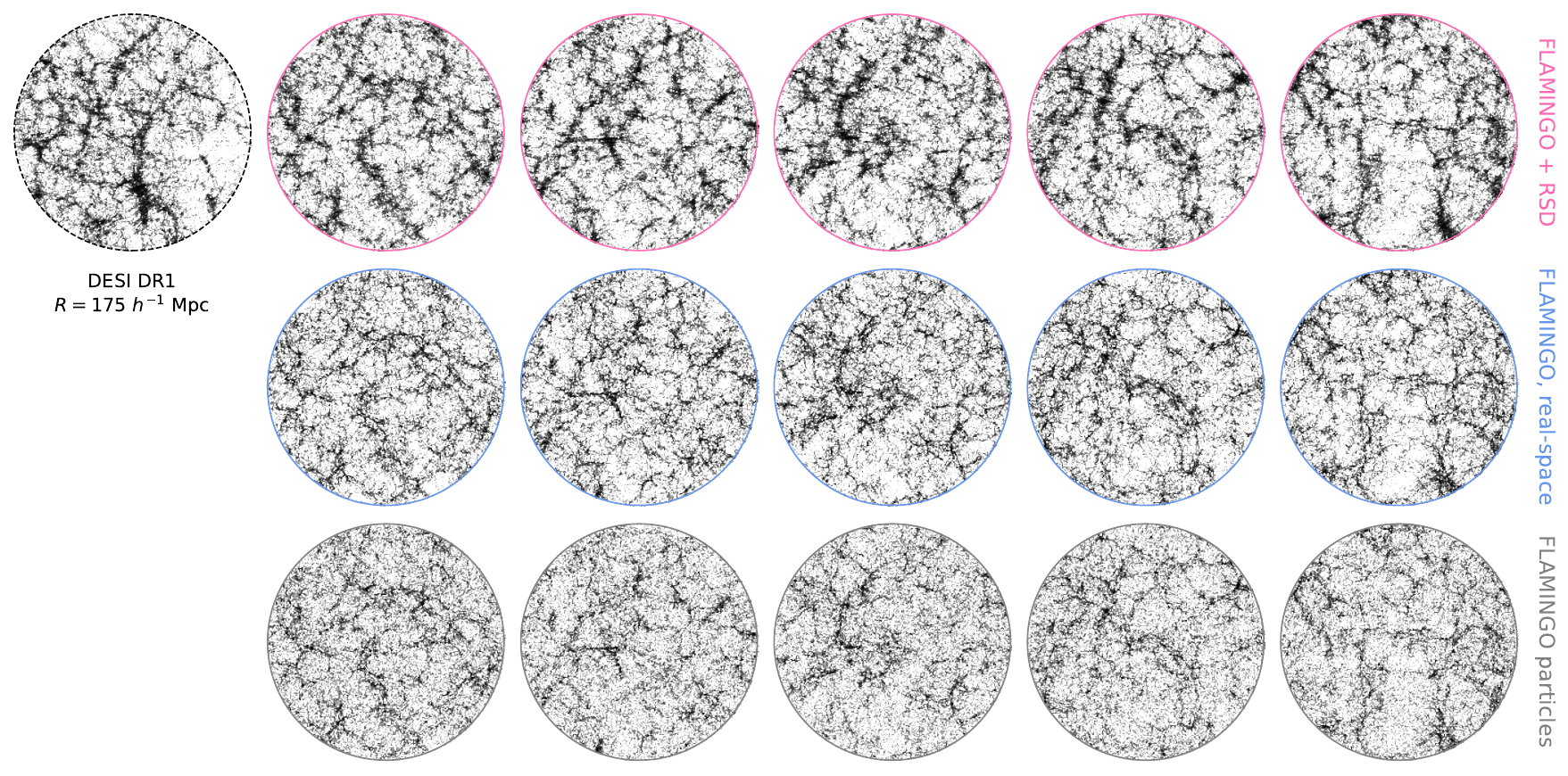}
\caption{\label{fig:structure-comparison-small} Distribution of galaxies within an $R = 175\,h^{-1}\,\mathrm{Mpc}$ cylinder from DESI DR1 (top left), and in five non-overlapping geometry-matched cylinders selected from FLAMINGO at $z=0.15$. The extent of the region approximates the true extent of the \texttt{S2} data. In the top row, the next five panels show FLAMINGO galaxies including redshift-space distortions. The middle row shows FLAMINGO galaxies in real space without redshift-space distortions. The bottom row shows the FLAMINGO density field sampled by dark-matter particles. Also at this scale, the amount of structure in DESI DR1 is visually similar to that in FLAMINGO galaxy catalogs with redshift-space distortions included. Without redshift-space distortions, the clustering is weaker, and the clustering in the particle distribution is the weakest.}
\end{figure*}

Figure~\ref{fig:structure-comparison-small} further demonstrates the consequence of missing galaxy- and halo bias, and missing redshift-space distortions. It compares the structures in DESI~DR1 data to FLAMINGO data in a cylinder of radius $R=175\,h^{-1}{\rm Mpc}$, which approximates the true extent, in comoving coordinates, of the \texttt{S2} region within the DESI data. As in Figure~\ref{fig:structure-comparison-large}, the simulation data are shown using either dark matter particles, galaxies in real space, or galaxies in redshift space. Also viewed on this scale, the FLAMINGO simulation matches the observed clustering strength, but only when galaxy catalogs including redshift-space distortions are used. Power spectra for this comparison are also shown in Figure~\ref{fig:DESI-power-spectra-individual}.

\section{Windowed power spectra} \label{sec:methods:PS}
The clustering strength within the cylinders is quantified using windowed two-dimensional power spectra in circular apertures. The same estimator is applied to DESI, \texttt{S2}, and all FLAMINGO mocks. This is important because the finite circular aperture has a non-trivial window function. No attempt is made to deconvolve the windowed spectra, instead they are directly compared to each other.

After projection into the $xy$ plane (where necessary, the \texttt{S2} data are already 2D), the data fall within a circular aperture of radius $R$, which is embedded within a square of side length $L_{\rm box}=2R$. The square is divided into $N_{\rm grid}=512$ cells per dimension. The overdensity is then computed as 
$$
\delta_{ij}= N_{ij}/\bar N - 1,
$$
where $N_{ij}$ is the number of galaxies within a cell (zero outside $R$ by construction), and the average $\bar N$ is taken only over cells within $R$. For cells outside $R$, $\delta = 0$.

To measure the power spectrum of the windowed density field, the discrete Fourier transform
$$
\tilde{\delta}(\boldsymbol k)
=
\Delta x^2\,{\rm FFT}(\delta_{ij})
$$
is computed, which leads to the two-dimensional power estimates for each mode
$$
P_{\rm 2D}(\boldsymbol k)
=
\frac{|\tilde{\delta}(\boldsymbol k)|^2}{L_{\rm box}^2}.
$$
Finally, $P_{\rm 2D}(\boldsymbol k)$ is averaged in circular annuli in Fourier space to obtain $P_{\rm 2D}(k)$.

\begin{figure*}
\centering 
\includegraphics[width=7.3cm, trim={0.0cm 1.2cm 0.0cm 0.0cm},clip]{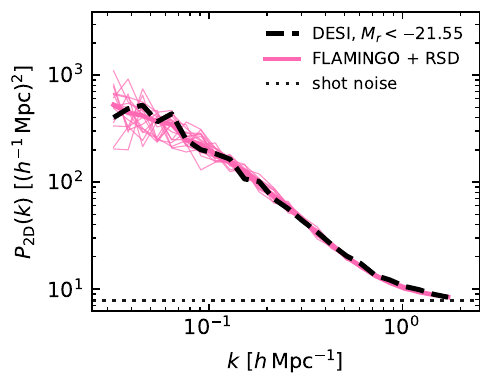}
\includegraphics[width=7.3cm, trim={0.0cm 1.2cm 0.0cm 0.0cm},clip]{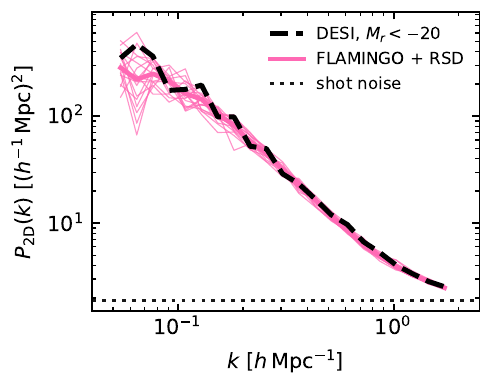}

\includegraphics[width=7.3cm, trim={0.0cm 1.2cm 0.0cm 0.0cm},clip]{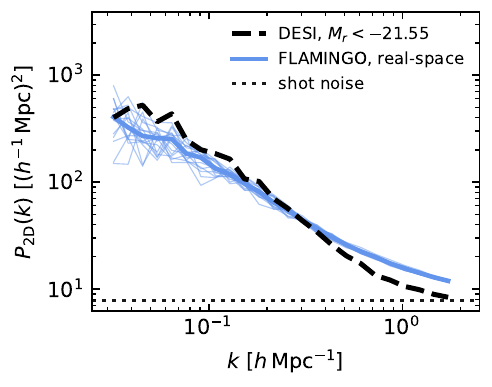}
\includegraphics[width=7.3cm, trim={0.0cm 1.2cm 0.0cm 0.0cm},clip]{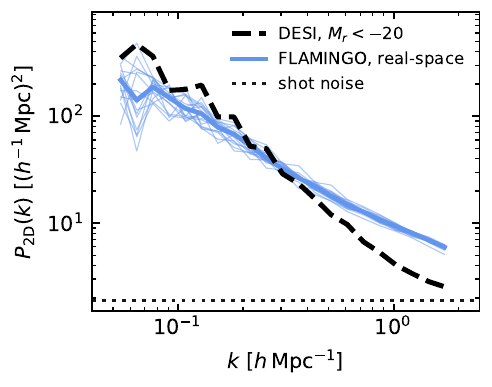}

\includegraphics[width=7.3cm, trim={0.0cm 0cm 0.0cm 0.0cm},clip]{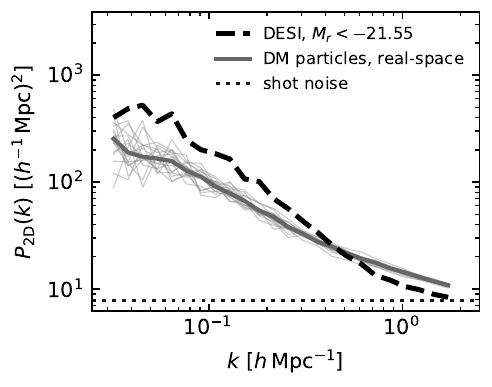}
\includegraphics[width=7.3cm, trim={0.0cm 0cm 0.0cm 0.0cm},clip]{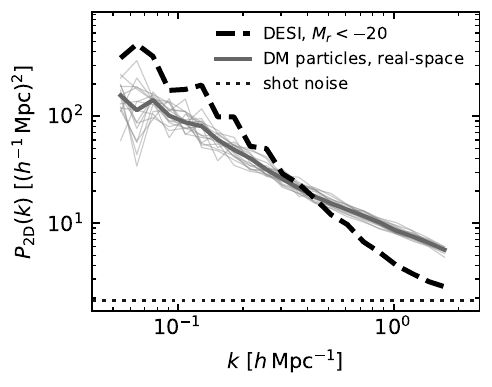}

\caption{\label{fig:DESI-power-spectra-individual}Power spectra comparison between DESI and individual slices of FLAMINGO at $z=0.15$. The left column shows the comparison in cylinders of $R = 290\,h^{-1}\,\mathrm{Mpc}$, similar to Figure~\ref{fig:DESI-power-spectra}. The right column shows the comparison in cylinders of $R = 175\,h^{-1}\,\mathrm{Mpc}$, approximately the true size of the \texttt{S2} region. Thin colored lines show individual power spectra in slices of the FLAMINGO simulation, from galaxy positions in redshift space (top), galaxy positions in real space (middle) or simulation particles (bottom). Thick lines show the corresponding medians. Dotted lines indicate the shot-noise level. When redshift-space distortions are included, and the comparison is performed at the correct scale, the DESI DR1 data are fully compatible with the FLAMINGO data on all scales.
}
\end{figure*}

\begin{figure*}
\centering 
\includegraphics[width=8.5cm, trim={0.0cm 0.8cm 0.0cm 0.0cm},clip]{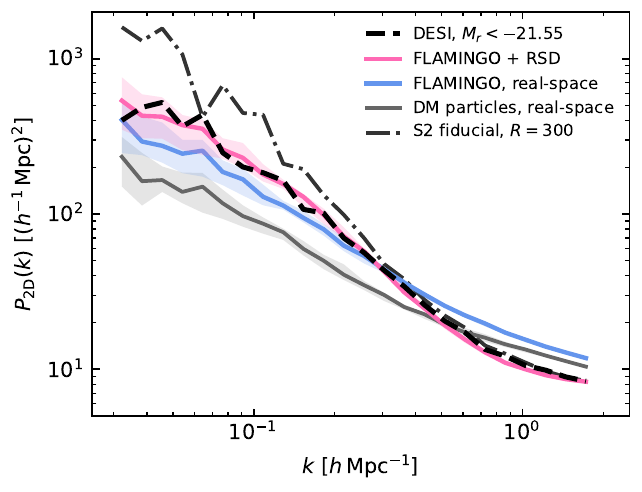} \\
\includegraphics[width=8.5cm, trim={0.0cm 0.8cm 0.0cm 0.0cm},clip]{pk2d_summary_DESI_vs_FLAMINGO_large_with_S2_no_shotnoise_S2density_snap_075.pdf} \\
\includegraphics[width=8.5cm, trim={0.0cm 0.0cm 0.0cm 0.0cm},clip]{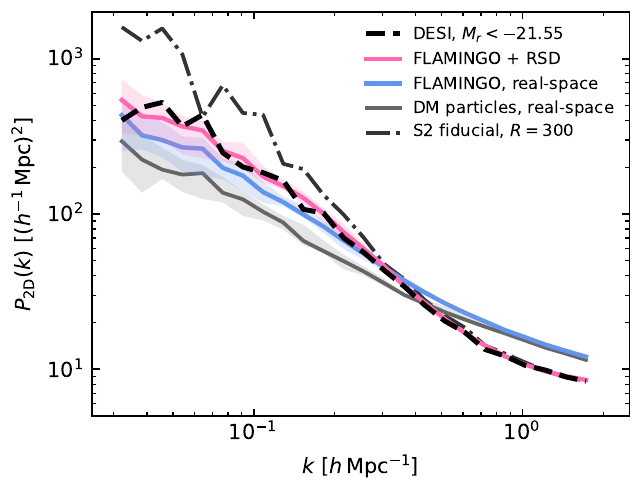}
\caption{\label{fig:DESI-power-spectra-z}Power spectra comparison between DESI, \texttt{S2}, and slices of FLAMINGO at different redshifts. As in Figure~\ref{fig:DESI-power-spectra}, on all panels, the dashed black line shows the power spectrum in the $R = 290\,h^{-1}\,\mathrm{Mpc}$ cylinder from DESI DR1, while the dash-dotted black line shows the power spectrum measured for \texttt{S2} in fiducial coordinates. From top to bottom, FLAMINGO data are computed from the snapshots at $z=0.25$, $z=0.15$ and $z=0.05$, with results at $z=0.15$ identical to Figure~\ref{fig:DESI-power-spectra}. The agreement with the DESI data and the disagreement with the \texttt{S2} data are unaffected by the choice of simulation output in this redshift range.
}
\end{figure*}

For the FLAMINGO comparison, the power spectra are measured independently in each non-overlapping cylinder. Figure~\ref{fig:DESI-power-spectra-individual} shows the equivalent of Figure~\ref{fig:DESI-power-spectra}, but for all individual cylinders. In Figure~\ref{fig:DESI-power-spectra-z}, the power spectra obtained from the snapshot at $z=0.15$, closest to the redshift of the center of the  $R=290\,h^{-1}\,\mathrm{Mpc}$ region, are compared to those obtained from simulation snapshots at $z=0.05$ and $z=0.25$ in the simulation, closest to the redshifts of the nearest and farthest data points. All three results agree closely with the DESI DR1 result.

\section{Distribution of angular pairwise distances} \label{sec:methods:adpd}
The distribution of angular pairwise distances measures the degree of anisotropy by measuring the clustering of orientations of pairs of points (such as the positions of galaxies or dark matter particles) as a function of the pairwise separations. The scale-dependent anisotropy is then quantified as the amount of clustering, or variance, of the angles at a given separation length. 

The points within each cylindrical volume are first projected onto the
two-dimensional \(xy\) plane. For every unordered pair of points \(i<j\),
the projected coordinate differences are defined as
\begin{equation*}
    \Delta x_{ij} = x_j-x_i,
    \qquad
    \Delta y_{ij} = y_j-y_i.
\end{equation*}
The projected pair separation and orientation are then
\begin{equation*}
    r_{ij}
    =
    \sqrt{\Delta x_{ij}^{2}+\Delta y_{ij}^{2}},
\end{equation*}
and
\begin{equation*}
    \theta_{ij}
    =
    \operatorname{atan2}(\Delta y_{ij},\Delta x_{ij})
    \bmod \pi,
\end{equation*}
respectively. Here, \(\theta_{ij}\) is the orientation of the undirected
line joining the pair relative to the \(x\)-axis and is folded into the
interval \(0\leq\theta<\pi\), or equivalently
\(0^\circ\leq\theta<180^\circ\). From this, a binned 2D density distribution is computed as: 

\begin{equation}
    p(\theta_k,r_j) = \frac{n_{jk}}{N_j},
\end{equation}

where \(n_{jk}\) is the number of pairs in radial bin \(j\) and angular bin \(k\), and $N_j = \sum_{k=1}^{N_\theta} n_{jk} $ is the total number of pairs in radial bin \(j\). At a given radial scale, $r_j$, the degree of anisotropy can then be measured as the variance, summing over the angular bins,

\[
\sigma_\theta^2(r_j)=\frac{1}{N_\theta}\sum_{k=1}^{N_\theta}
\left[p(\theta_k,r_j)-\frac{1}{N_\theta}\right]^2,
\]

where the normalizing factors of $\frac{1}{N_\theta}$ correspond to the occupation of the $N_\theta$ angular bins for an isotropic distribution. The practical implementation is taken directly from the code published by \cite{SylosLabini-2026d}.

\begin{figure*}
\centering 
\includegraphics[width=8.5cm, trim={0.0cm 0.0cm 0.0cm 0.0cm},clip]{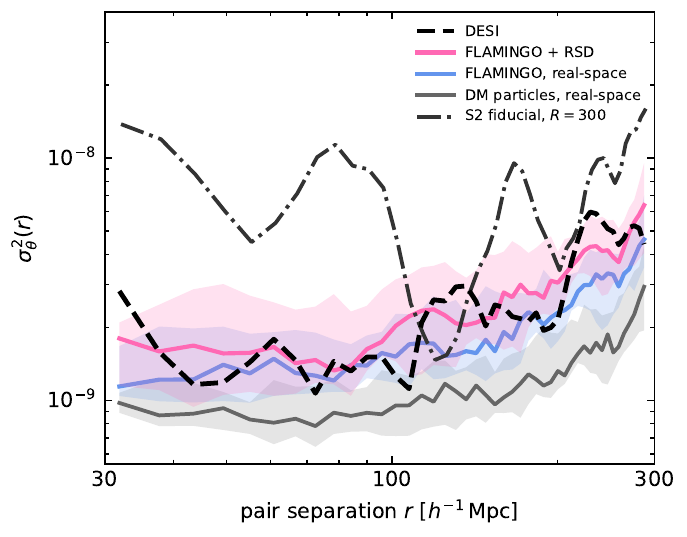}

\caption{\label{fig:angular-PS}Anisotropy as a function of scale, measured from the distribution of angular pairwise distances, for the DESI DR1 data, the FLAMINGO simulation data, and the \texttt{S2} data in its fiducial coordinates. The anisotropy in the FLAMINGO galaxy catalogs that include redshift-space distortions matches that of the DESI data, when compared in the correct coordinates. The absence of redshift-space distortions (in the real-space galaxy catalogs), or the absence of both redshift-space distortions and bias (in the particle data) reduces the anisotropy in the simulation. By contrast, the \texttt{S2} data show spuriously high anisotropy.
}
\end{figure*}

In Figure~\ref{fig:angular-PS}, we show the variance, $\sigma_\theta^2(r)$, measured in the DESI data, the FLAMINGO simulation data, and the \texttt{S2} data in its fiducial coordinates.

\newpage

\clearpage

\bibliographystyle{aasjournalv7}
\bibliography{bibliography}

\end{document}